\documentclass[journal,twoside,web]{IEEEtran}
\usepackage[utf8]{inputenc}
\usepackage{cite}
\usepackage{amsmath,amssymb,amsfonts}
\usepackage{graphicx}
\usepackage{algorithm,algorithmic}
\usepackage{textcomp}
\usepackage{multirow}
\usepackage{booktabs}
\usepackage{subcaption}
\usepackage{caption}
\usepackage{float}
\usepackage{comment}
\usepackage[pdfstartview=XYZ,
bookmarks=true,
colorlinks=true,
linkcolor=blue,
urlcolor=blue,
citecolor=blue,
pdftex,
bookmarks=true,
linktocpage=true,
hyperindex=true
]{hyperref}
\usepackage{array}
\usepackage{hyperref}

\def\rd{black}

\newcolumntype{C}[1]{>{\centering\arraybackslash}p{#1}}

\usepackage{orcidlink}

\def\BibTeX{{\rm B\kern-.05em{\sc i\kern-.025em b}\kern-.08em
    T\kern-.1667em\lower.7ex\hbox{E}\kern-.125emX}}
\begin{document}

\title{Simultaneous Control of Human Hand Joint Positions and Grip Force via HD-EMG and Deep Learning}

\author{Farnaz Rahimi \orcidlink{0009-0001-0284-5221}, Mohammad Ali Badamchizadeh \orcidlink{0000-0002-9999-1152}, Raul C. Sîmpetru \orcidlink{0000-0003-0455-0168}%\IEEEmembership{Graduate Student Member, IEEE}
, Sehraneh Ghaemi \orcidlink{0000-0002-4731-6577} 
, Bjoern M. Eskofier \orcidlink{0000-0002-0417-0336}%\IEEEmembership{Senior Member, IEEE} 
, and Alessandro Del Vecchio \orcidlink{0000-0002-7118-5554}% \IEEEmembership{Member, IEEE}
\thanks{Farnaz Rahimi and Bjoern Eskofier received funding from the Deutsche Forschungsgemeinschaft (DFG, German Research Foundation) – SFB 1483 – Project-ID 442419336, EmpkinS. The work of Alessandro Del Vecchio was supported by the European Research Council (ERC) Starting Grant project GRASPAGAIN under grant 101118089 and by the German Ministry for Education and Research (BMBF) through the project MYOREHAB under Grant 01DN2300.
}
\thanks{ Farnaz Rahimi, Mohammad Ali Badamchizadeh and Sehraneh Ghaemi are with the Faculty of Electrical and Computer
Engineering, University of Tabriz, Tabriz 5165677861, Iran (e-mail: f.rahimi@tabrizu.ac.ir, mbadamchi@tabrizu.ac.ir, ghaemi@tabrizu.ac.ir).}
\thanks{Raul C. Sîmpetru, Bjoern M. Eskofier and Alessandro Del Vecchio are with the Department Artificial Intelligence in Biomedical Engineering, Friedrich-Alexander-University Erlangen-Nuremberg, Erlangen, 91052, Germany (e-mail: raul.simpetru@fau.de, bjoern.eskofier@fau.de, alessandro.del.vecchio@fau.de ).}}

\maketitle
\begin{abstract}
In myoelectric control, simultaneous control of multiple degrees of freedom can be challenging due to the dexterity of the human hand. Numerous studies have focused on hand functionality, however, they only focused on a few degrees of freedom. In this paper, a 3DCNN-MLP model is proposed that uses high-density sEMG signals to estimate 20 hand joint positions and grip force simultaneously. The deep learning model maps the muscle activity to the hand kinematics and kinetics. The proposed models' performance is also evaluated in estimating grip forces with real-time resolution. This paper investigated three individual dynamic hand movements (2pinch, 3pinch, and fist closing and opening) while applying forces in 10\% and 30\% of the maximum voluntary contraction (MVC). The results demonstrated significant accuracy in estimating kinetics and kinematics. \textcolor{\rd}{The average Euclidean distance across all joints and subjects was 11.01 ± 2.22 mm and the mean absolute error for offline and real-time force estimation were found to be 0.8 ± 0.33 N and 2.09 ± 0.9 N respectively.} The results demonstrate that by leveraging high-density sEMG and deep learning, it is possible to estimate human hand dynamics (kinematics and kinetics), which is a step forward to practical prosthetic hands.

\end{abstract}


\section{Introduction}
\label{sec:introduction}
\begin{figure*}[htbp]
  \centering
  \includegraphics[width=\textwidth]{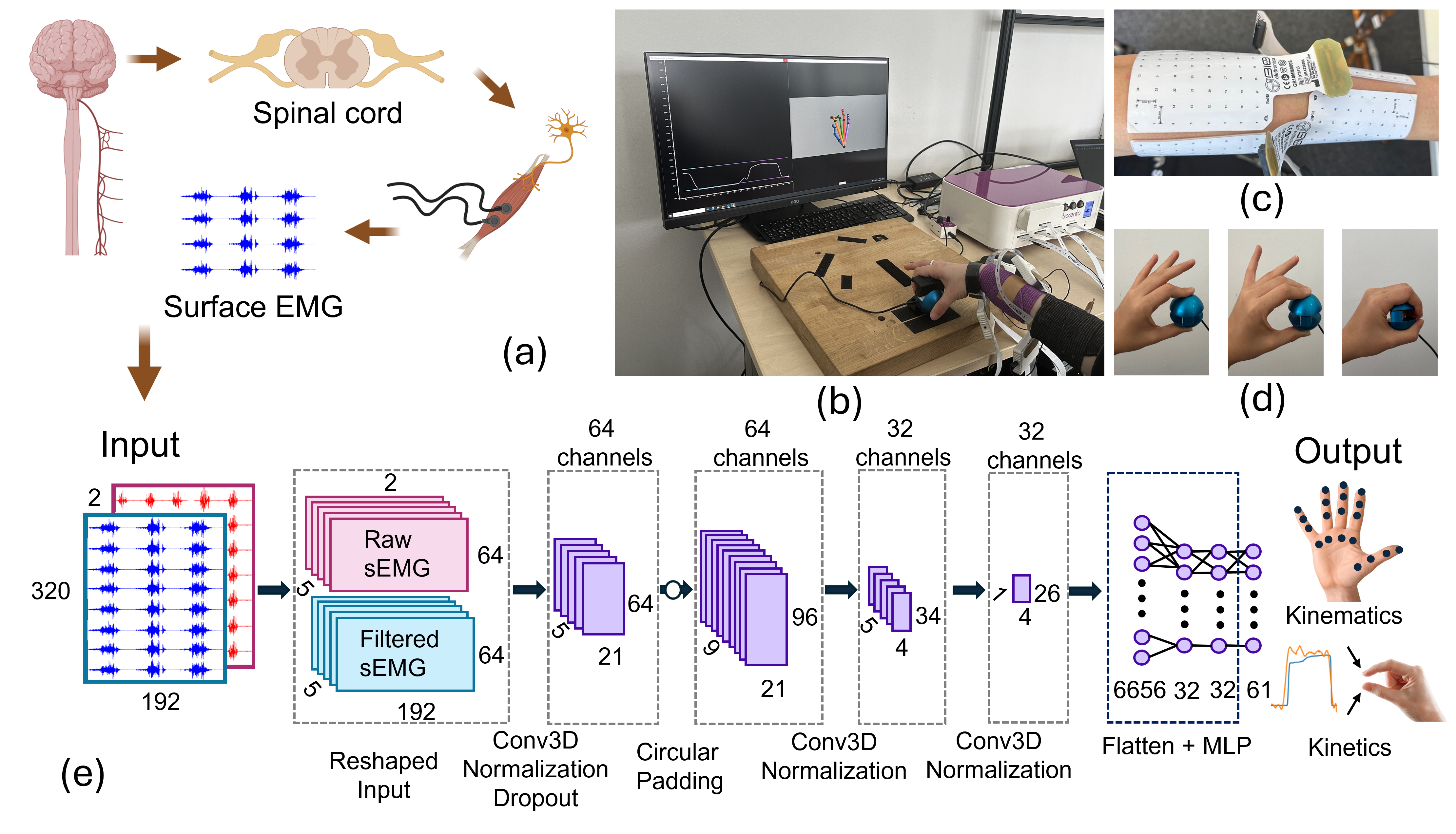}
  \caption{Overview of the study. 
    \hyperref[fig:a]{(a)} Neuromuscular pathway showing the flow of signals from the brain through the spinal cord to muscles.
    \hyperref[fig:b]{(b)} Experiment setup. The participant is following the hand videos while reaching the predefined force level.
    \hyperref[fig:c]{(c)} Three EMG grids are placed around the arm under the elbow and two grids proximal to the distal ulnar head.
    \hyperref[fig:d]{(d)} Three hand movements performed during the experiment, 2pinch, 3pinch and grasp. 
    \hyperref[fig:e]{(e)} Deep learning model architecture for predicting hand joint positions and forces. }
  \label{fig: overview}
  \label{fig:a}
  \label{fig:b}
  \label{fig:c}
  \label{fig:d}
  \label{fig:e}
\end{figure*}

\IEEEPARstart{T}{he} human hand is a complex and highly functional part of the body, doing a wide range of movements in daily life activities. It comprises multiple joints, leading to its dexterity. Accurate and precise controlling of these multiple degrees of freedom (DoF) is essential for developing prosthetic devices. \textcolor{\rd}{Biosignals are extensively employed in human-machine Interfaces (HMIs) \cite{JAfari},\cite{Yang}. Among these signals, Electromyography (EMG) is prominently used for controlling myoelectric devices.}
Surface EMG (sEMG) \cite{merletti} is a non-invasive method for capturing the electrical activity generated by ensembles of motor units (MUs) from the surface of the skin (Fig \hyperref[fig: overview]{ \ref*{fig: overview}a}). It contains valuable information on muscle contractions, therefore making it a suitable option for controlling rehabilitation systems, like exoskeletons and prosthetics. %\cite{Batayneh}. 
Recent developments in myoelectric control \cite{Chen, Simpetru2, Wei} 
facilitated the prediction of kinematics and kinetics, but remain limited in their ability to fully control hand functions as it is still challenging to decode the joint forces into control signals. Most of the studies primarily focused on estimating discrete hand gesture recognition \cite{Wei}, \cite{Shen}, grip forces \cite{Fang}, and joint angles \cite{Olsson} from sEMG patterns, therefore a limited number of DoFs could be controlled, resulting in restricted motion output. 

Natural hand movements are not limited to discrete patterns. Continuous estimation and adjustment of the control signals lead to smoother and more natural movements of prosthetics. Therefore continuous controlled strategies are preferred to discrete classification-based control 
\cite{Ngeo, Ding, Pan}.

Research on myoelectric control has mainly focused on either kinematics or kinetics in isolation\cite{Martinez,Zhao,Zhao2023,Dai,Zhang }. Various algorithms %of regression and \textcolor{\rd}{deep} learning 
have been developed for estimating isometric finger or grip forces \cite{Jiang, Zheng, Xue, Martinez }. Similar algorithms have been used, focusing mainly on kinematics, for instance, wrist joint angle estimation \cite{Zhao}, \cite{Zhao2023}, and finger joint angles \cite{Dai}, \cite{Zhang}. However, for more effective myoelectric control, it is crucial to address both kinetics and kinematics simultaneously. 

For the human hand, the wrist and finger motions together with the forces exerted, largely determine its functionality. Therefore recent studies have focused on the simultaneous estimation of wrist/finger motion and forces \cite{Li, Mao, Roy,Rahimi, Sun} to have a better understanding of the human hand.  
The aforementioned studies have good results in simultaneous estimation, however, they only evaluated a few DoFs. For wrist motion, three DoFs were studied including wrist flexion/extension, abduction/adduction, and pronation/supination. Considering grip force in total 4 DoFs have been estimated in \cite{Li} and \cite{Mao}. Finger angles and forces are studied in \cite{Roy} considering the same number of DoFs. In our previous study \cite{Rahimi} four hand gesture types and five individual finger forces were estimated simultaneously considering 6 DoFs in total. Sun et al.\cite{Sun} estimated the finger curvatures in five DoFs and grip force in one DoF. However, the human hand has a complex anatomy. It has 5 digits and 15 joints. 
Simultaneous estimation of multiple DoFs from sEMG is challenging due to the complexity of modeling the large input-output space. Additionally, multiple muscles from superficial to deep are involved in hand movements and there are inter-dependencies between finger movements, which means that the movement of a single finger or a single DoF within the same finger cannot be performed completely independently \cite{Ingram}. %\cite{Batayneh}
Therefore only a few studies have focused on replicating the multi-DoF movements of the human hand.
Numerous studies have successfully predicted continuous joint angles but with a smaller number of DoFs \cite{Chen, Guo, Serdana}. Recent advancements in machine learning have paved the way for studying all hand joints to reconstruct the full human hand. Previous studies \cite{Batayneh},  \cite{Liu} utilized different artificial neural networks to continuously estimate hand joint angles from the sEMG signals. Other studies \cite{Quivira}, \cite{Simpetru} focused on analyzing hand movements by reconstructing the full human hand kinematics as 3D points in Euclidean space. These studies have good results for kinematics but they have overlooked the forces. The estimation of both kinetics and kinematics with a high number of DoFs is not well addressed in the literature which is essential for a practical and intuitive prosthetic hand. 
Considering the lack of research in multi-DoF kinetics and kinematics estimation, in this paper, a deep learning model is proposed to construct the full human hand. The model aims to predict 20 joint positions along with the grip force representing the complete movement and force dynamics of the hand.

This paper further investigates the real-time estimation of grip forces, focusing on the kinetics of the human hand to enhance the effectiveness of force control in prosthetic devices. 

The main contribution of this paper is to address the following key points within the state of art:
\begin{itemize}
    \item 
    
    \textbf{Multi-DoF kinetics and kinematics estimation}
    
 Considering the lack of research on simultaneous kinetics and kinematics estimation, this paper aims to estimate both hand joint movements and exerted grip forces. There are only a few studies \cite{Li, Mao, Roy, Rahimi} on this issue limited to a few DoFs for wrist motion and grip force. In this paper, we took a distinct path by estimating 20 human hand joint positions and grip forces (21 DoFs) over 3 movements.
 \item
 
 \textbf{Real-time grip force estimation}
 
 The efficiency of the deep learning model in kinematics estimation is investigated in \cite{Simpetru2}. In the current paper, grip forces are estimated with real-time resolution to evaluate the performance of the deep learning model in kinetics estimation.
\end{itemize}

\section{Methods}

\subsection{Data acquisition}
The data acquisition setup is shown in Fig. \hyperref[fig: overview]{ \ref*{fig: overview}b}. Nine healthy right-handed individuals participated in the study, comprising four men and five women aged between 23 and 30 years. They signed an informed consent before the experiments. The experiments were in agreement with the Declaration of Helsinki and approved by the Friedrich-Alexander University ethics committee (n. 21-150 3-B). The participants were asked to shave their forearms before the experiment and clean the skin with alcohol. Five sEMG grids \textcolor{\rd}{each measuring} 8 rows by 8 columns with an interelectrode distance of 10 mm (OT Bioelettronica, Turin, Italy) were \textcolor{\rd}{attached} around the forearm and wrist \textcolor{\rd}{(Fig \hyperref[fig: overview]{ \ref*{fig: overview}c})} \cite{SimpetruInfluence}. \textcolor{\rd}{ The self-adhesive bandages were used to firmly secure the electrode grids and ensure they remained fixed during the testing. 
The analog HD-sEMG signals were recorded in monopolar mode with a 150× amplification. The sampling frequency was 2048 Hz, signals were bandpass filtered between 10–500 Hz and converted to digital format using a multichannel amplifier with a 16-bit analog-to-digital converter (EMG-Quattrocento, OT Bioelettronica, Turin, Italy).} The grip force was recorded using a force dynamometer (COR2, OT Bioelettronica, Turin, Italy) connected to EMG-Quattrocento. The force dynamometer was calibrated prior to the experiments with different weights to obtain the force values in Newtons. The HD-sEMG and the grip force were recorded simultaneously and synchronized directly at the source. The participants sat in front of a desk with a screen in front of them to follow the video of the movements. The force dynamometer was fixed on the desk, and the participants had to press it to follow the predefined force trajectories. % (Figure \hyperref[fig: overview]{ \ref*{fig: overview}b}).

In order to have a generalized model and avoid restricting it to a specific hand skeleton, the ground truth kinematics data were recorded from a single individual, then it was used to guide the other individuals to mimic the same movements during the experiments. Four cameras simultaneously recorded the movements from four different angles. \textcolor{\rd}{The videos were first processed with Deeplabcut \cite{Mathis}, a markerless kinematics software, and then aligned in 3D space using Anipose \cite{Karashchuk}.} Based on the recorded kinematic data for one subject, a video was created for each movement. The videos were displayed to the participants, and they were asked to follow them to have a consistent frequency for repeating the movements (detailed explanation in Cakici et al.\cite{ Cakici}). The kinematics of one subject were used as a ground truth for all participants. This made it possible to skip the kinematics recording for each participant and allowed subjects who could not move their hands to take part in the study in the future.
 The kinematics data in this paper were previously used in \cite{Simpetru2} but from a different perspective. In the previous study, the hand exercises were performed vertically. However, in this paper, the hand was aligned horizontally to the ground to grip the force dynamometer, so the kinematics were rotated by 90 degrees to match this new alignment. 

 The participants repeated each movement for 60 seconds. \textcolor{\rd}{The hand exercises were 2pinch, 3pinch, and the fist (Fig \hyperref[fig: overview]{ \ref*{fig: overview}d}). The participants performed each exercise twice, applying 10\% and 30\% of their Maximum Voluntary Contraction (MVC). }
They were instructed to follow the hand kinematics video, press the force dynamometer to reach the specified force levels (10\% or 30\% MVC), and then release it. Real-time feedback was provided to show the force level changes while performing the movements.

\subsection{Preprocessing}
The collected data required preprocessing before being fed to the network. The sEMG and force data were recorded in non-overlapping windows of 64 samples providing a steady rate of 32 predictions per second. Simultaneously, the timestamps corresponding to each segment in the video frames were calculated. After recording, these intervals were used to save the synchronized kinematics and EMG data. According to \cite{Simpetru2}, the windows of 64 samples of sEMG did not provide enough temporal information, therefore 3 consecutive windows were combined to form a longer 192-sample window with an increment of 64 samples. The kinematics and forces do not vary significantly in 94 ms (192 samples). Therefore, the average of the windows was considered as model output. This simplified the model’s output from a matrix to a vector for each sEMG window. Each window of the sEMG signal was low-pass filtered ($<$ 20 Hz) with a 4th-order digital Butterworth filter and appended to the raw windows. Low-pass filtered sEMG has been proven to improve the deep learning models' performance \cite{Simpetru},\cite{Simpetru2022}.

\subsection{EMG augmentation}
Deep learning models require a large amount of data for training. Therefore, the collected data was augmented using three different augmentation methods. %Augmentation helps the model learn different data architectures and results in robustness. 
In this paper, three augmentation methods were used. First, Gaussian noise \cite{Tsinganos} was added to each sEMG channel to have a signal-to-noise ratio of 5. This helps the model to learn realistic noisy conditions. The other augmentation method was the Magnitude wrapping technique \cite{Tsinganos}. Each sEMG channel was multiplied by a curve sampled from a normal distribution. The wrapped signal had the same characteristics as the original sEMG signal, but the amplitude varied according to the wrapping curve. This slight variation simulated the potential electrode displacement during the experiments. Finally, the Wavelet decomposition \cite{Tsinganos} was applied to reconstruct distorted sEMG signals. These data augmentation methods resulted in a fourfold increase in data and prepared the model for real-life conditions and challenges. 

\begin{figure*}[htbp]
%\vspace{-\baselineskip}
  \centering
  \includegraphics[width=0.95\textwidth]{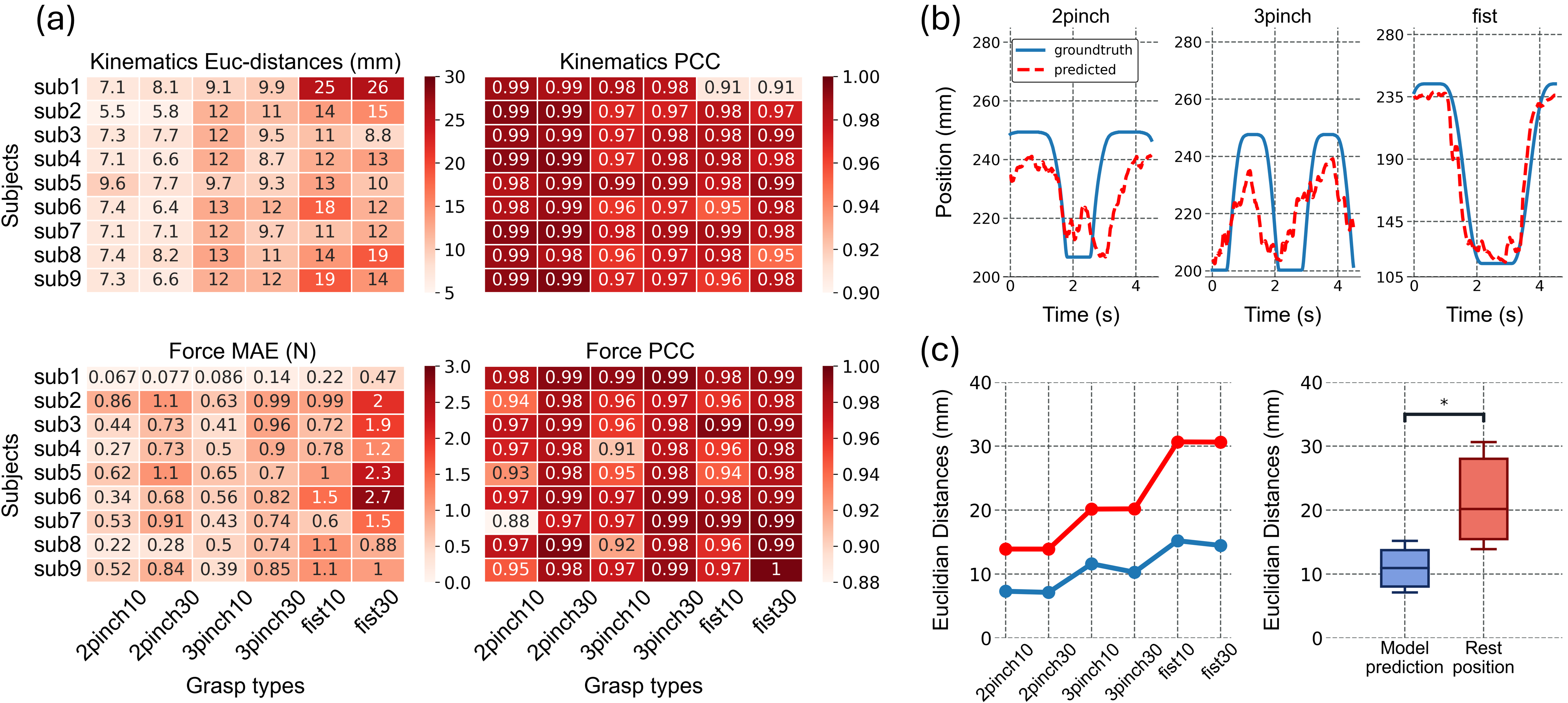}
  \caption{The proposed model performance. 
  \hyperref[fig:a]{(a)} Heatmaps of error metrics for kinematics and force estimation across all subjects and postures: Kinematic errors represented by Euclidean Distance and PCC, averaged over 20 joints and 3 axes; Force estimation errors represented by MAE and PCC (10 and 30 represent the relative force level).
  \hyperref[fig:b]{(b)} Index fingertip movement prediction across different postures for subject 3 (joint no. 5).
  \hyperref[fig:c]{(c)} The proposed model's kinematics prediction performance compared to the steady prediction of rest position. The kinematics prediction accuracy is significantly different from the steady rest position (Student's t-test, * p $<$ 0.05). 
  }
 % \vspace{-\baselineskip}
  \label{fig: heatmap}
\end{figure*}

\subsection{Model}
The model employed in this paper (Fig \hyperref[fig: overview]{ \ref*{fig: overview}e}) is an adaptation of one from our earlier paper \cite{Simpetru2}. The prior model was outputting either kinematics or kinetics. The main contribution of this paper is to have a simultaneous estimation of both kinetics and kinematics. To achieve this, the model is modified to output all hand joint positions in 3D space as well as the grip force simultaneously. The input tensor is the combined raw and low-pass filtered sEMG windows with the shape of 2 (raw and filtered sEMG inputs) × 320 (number of channels) × 192 (number of samples per window). To apply grid-wise normalization \cite{Simpetru2}, the channels are split by the number of grids into a 4D tensor of shape 2 × 5 (number of sEMG grids) × 64 (number of electrodes per grid) × 192. The model begins with a 3D \textcolor{\rd}{convolutional} layer (kernel size = 1 × 1 × 31, stride = 1 × 1 × 8, number of outputs = 64). This layer searches the individual channels in a window of 31 samples (15.2 ms) to find the action potentials of the sEMG signals. The activation function used for the convolutional layers is GELU \cite{Hendrycks}, and each convolutional layer is followed by an Instance Normalization layer \cite{Ulyanov}. Then there is a 3D dropout layer (p = 0.25) to prevent overfitting. The next layer is a circular padding (2 for back and front; 16 for top and bottom) which extends the extracted features dimension and preserves the recording's synchronicity for the subsequent layer. The second 3D convolutional layer has a kernel size of 5 × 32 × 18 with dilation of 1 × 2 × 1 and 32 output channels. This layer looks at all 5 grids at once and extracts the features of the grid combination. The last 3D convolutional layer has a kernel size of 5 × 9 × 1 with 32 output channels, to filter out the unnecessary information and collect the most relevant details. 
The extracted features from the previous convolutional layers are flattened and fed into a simple perceptron with three layers. The first two layers have 32 neurons with GELU activation functions, and the last layer is the output.

The model output is the 3D hand skeleton including all 20 joint values in 3 dimensions (60D) and a grip force (1D). The wrist position is constant and excluded from the output. The model is adapted to learn fewer tasks (2 pinch, 3 pinch, and fist closing and opening) by reducing the number of channels per layer. 
Mean absolute error is chosen as the loss function and AdamW \cite{Loshchilov} with the AMSGrad \cite{Reddi} correction is used for model optimization. The parameters used for training the model are fully explained in \cite{Simpetru2}.

\section{Real-time Interface}

Real-time control provides immediate feedback to achieve natural and intuitive control and improves performance and user engagement. This paper focuses mainly on real-time force estimation as kinematics has previously been studied in \cite{Simpetru2}. The real-time experiment was conducted on separate days with one or two-day intervals between sessions. After collecting data on the first session, the model was trained, and the offline results were calculated. The placement of the sEMG grids was slightly different in two separate sessions as it was impossible to attach them in the same position. Therefore, during the second session, each movement was repeated for 30 seconds, and the data were used to transfer learning. The trained model was retrained for 12 epochs with new data to be adapted to the new electrode positions preparing it for real-time predictions. 
The sEMG and force data were recorded in non-overlapping windows of 64 samples. As previously noted a window of 64 did not contain enough temporal information, in offline experiments, three windows of data were combined to create a window of 192 samples. In the real-time application, the model needed to wait for two additional sEMG windows to make useful predictions. This means that the model required a 93.75 ms warm-up time. After this warm-up period, the model generated 32 predictions per second after receiving each new sEMG window (31.25 ms). The model output was refined using the real-time filter introduced in \cite{Simpetru2} to remove the jitter and correct the predictions. This filter performed well in real-time experiments as it has no delays and uses the memory of previous predictions to reconstruct predictions accurately.

\section{Results}
The acquired data were split into training and testing sets with an 80-20 ratio. The training set was further divided into training and validation sets. The deep learning model was trained with the training set for 50 epochs. To evaluate the model's performance, Pearson correlation coefficient (PCC) and mean absolute error (MAE) were used as error metrics for force prediction results. The prediction error for kinematics was calculated in Euclidean distance in millimeters (mm) to represent the hand joint position error in 3D along with PCC. The model effectively predicted 20 hand joint positions along with the applied forces demonstrating its performance in simultaneous prediction of both kinetics and kinematics. The errors for each subject and task are summarized in Fig \hyperref[fig: heatmap]{ \ref*{fig: heatmap}a}.

\subsection{ Kinematics prediction performance}
The kinematics prediction errors averaged across all 20 joints, three dimensions \textcolor{\rd}{(X, Y, Z)}, and all time steps. The mean Euclidean distances and PCC across all subjects are 11.01 ± 2.22 mm and 0.98 ± 0.01 respectively. Figure \hyperref[fig: heatmap]{ \ref*{fig: heatmap}b} illustrates the Index fingertip movement for one of the subjects. The figure demonstrates how the Index fingertip’s position has changed along the Z axis during 2pinch, 3pinch, and fist movements. 

In order to further investigate the model’s performance in predicting kinematics, a statistical analysis (Student's t-test) is done to examine the absolute Euclidean distances across tasks under two different conditions. \textcolor{\rd}{The proposed model's prediction for 20 joints is compared to a baseline condition where the model has a steady output and predicts the rest state for all movements regardless of the task.} Figure \hyperref[fig: heatmap]{ \ref*{fig: heatmap}c} shows the mean Euclidean distance over all subjects for each task. Results indicate that the AI prediction is significantly different (p $<$ 0.05) from the rest state, demonstrating the model’s ability to accurately predict the 20 hand joint positions and consequently the hand movements.

\subsection{Force prediction performance - Offline}
According to Fig \ref{fig: heatmap} the deep learning model accurately predicted the forces across various hand movements and force levels. The applied forces correspond to 10\% and 30\% of the subject’s MVC. The mean MAE and PCC over all subjects are 0.8 ± 0.33 N and 0.97 ± 0.02 respectively. The results demonstrate that the model achieved reliable force predictions, and effectively interprets sEMG signals to predict forces with high accuracy and reliability. The force prediction results for subject 1 are shown in Fig \hyperref[fig:force_performance]{ \ref*{fig:force_performance}a}. 

\subsection{Force prediction performance - Real-time}
\begin{figure}[htbp]
  \centering
  \includegraphics[width=0.48\textwidth]{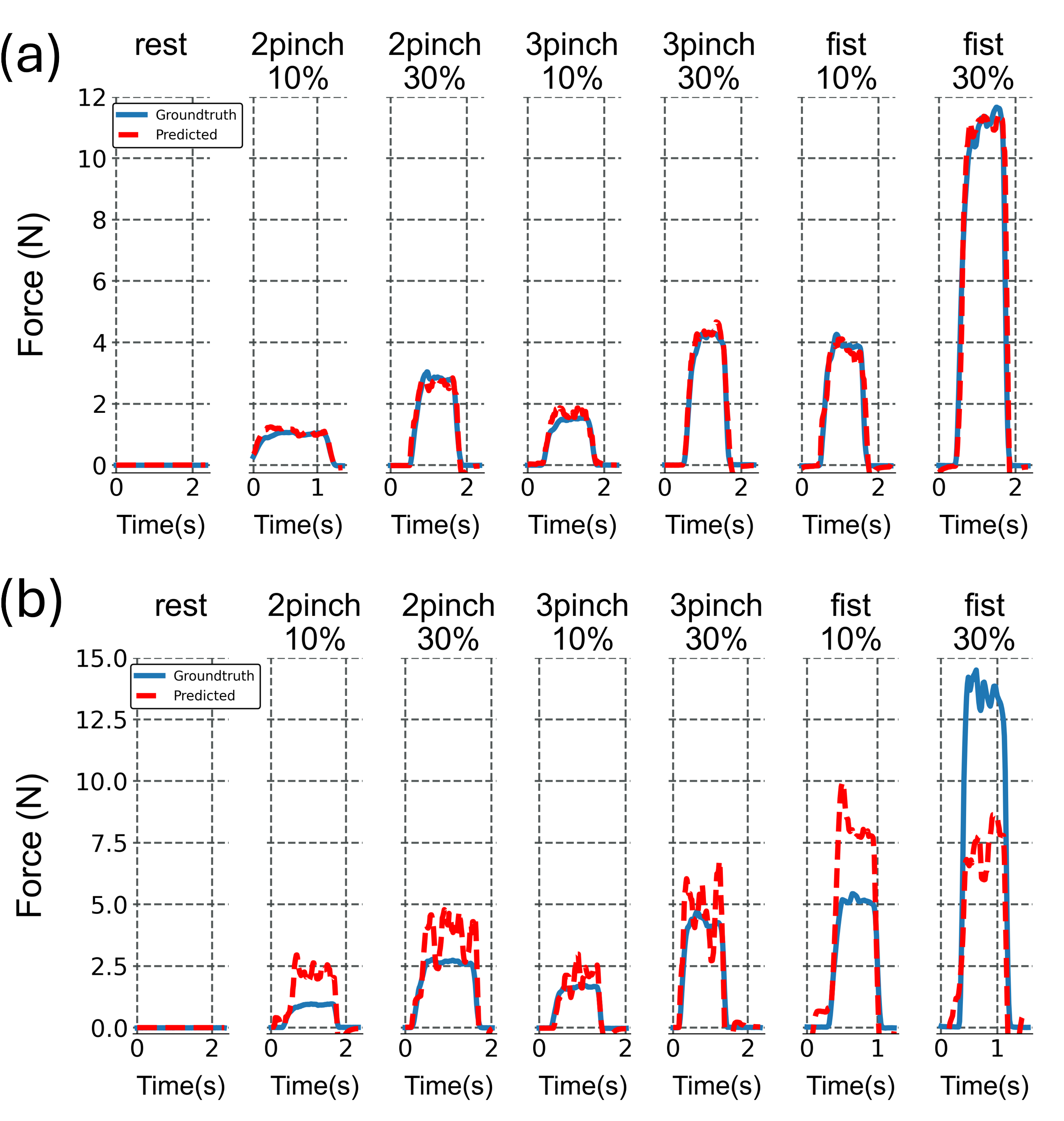}
  \caption{Force estimation results for subject 1 across various postures and force levels. The 10\% and 30\% values indicate force levels as a percentage of MVC.  \hyperref[fig:a]{(a)}  Offline force estimation results. \hyperref[fig:b]{(b)}  Real-time force estimation results. } 
  %\vspace{-\baselineskip}
  \label{fig:force_performance}
  \label{fig:a}
  \label{fig:b}
\end{figure}

This paper focuses on force estimation with real-time resolution to evaluate the model’s performance in real-time force estimation. 
The model’s performance for force estimation is evaluated through further experiments and analysis. After training the model with offline training data, a separate session was conducted for real-time experiments. Following the transfer learning process, the participants performed the same experiments with real-time feedback shown to them to have a proportional control on the applied and predicted forces. The model was outputting the force with 32 predictions per second after the warm-up period. The mean MAE and PCC over all subjects for real-time experiments are 2.09 ± 0.9 N and 0.92 ± 0.04 respectively. The model’s performance is relatively lower in the real-time experiments compared to offline analysis. This reduction in performance might be because of the shifts in electrodes as the real-time experiments were conducted several days after the offline experiments and it is challenging to place the sEMG grids in the exact same position as during the offline experiments. These shifts can lead to variations in signal acquisition affecting the model’s prediction. However, despite these challenges, the model was still able to accurately detect and predict forces demonstrating its robustness to different conditions. The real-time force prediction results for subject 1 are shown in Fig \hyperref[fig:force_performance]{ \ref*{fig:force_performance}b}.

\section{Discussion}

The main goal of this paper was to study the human hand focusing on the joints movements and forces exerted during movements. The human hand has an intricate structure with numerous joints. Extensive research has been conducted on the human hand to understand its functionality and replicate its movements. Most of the research focused on human hand discrete movements and obtained accurate performances. Recent studies moved toward continuous hand movement estimation as it provides a more accurate representation of natural movement. 
Additionally, to develop an accurate model of the hand it is essential to consider the forces involved in holding or gripping objects. The combination of kinetics and kinematics estimation is overlooked in the literature or restricted to limited DoFs (such as simultaneous estimation of wrist angle and forces (2 DoFs) or movement types and forces.). These approaches fail to capture the full functionality of the human hand. This paper aims to fill the existing gap by simultaneously estimating kinetics and kinematics across a higher number of DoFs ($> 20$). This method enables the exact positioning of each hand joint, capturing even the smallest movements and the forces applied.

\subsection{Comparison to other related works}

There are a few studies on simultaneous and continuous estimation of kinetics and the kinematics of the human hand. The most recent similar study \cite{Li} proposed a graph-driven method for simultaneous and proportional estimation of wrist joint angles and grip forces. Similar to our study, they used HD-EMG to capture the forearm muscle information and tried three different wrist movements. The participants followed the wrist trajectories and applied forces in 30\% and 60\% of their MVC. A new deep learning model based on graph CNN and LSTM is proposed to capture spatial and temporal information and estimate the wrist angle and the force exerted simultaneously. The proposed method's performance is compared with other deep learning models including LSTM, Conv-LSTM, and CNN. The graph CNN-LSTM achieved the best results among other models with the average PCC of 91.9\% for force and 89.6\% for wrist angle estimation over all subjects. 
In another similar study \cite{Mao}, wrist angle and grip force are studied during three wrist movements. Support vector Regression (SVR) is used to estimate both kinetics and kinematics. They have compared different feature sets such as the combination of EMG and Acceleration (ACC) signals. The best performance of the proposed method for force estimation using only EMG signals is 92.42 ± 0.87\% and using the combined EMG and ACC features they have achieved 95.32 ± 1.35 \% correlation. The wrist angle correlation using only EMG and EMG + ACC features are 84.41 ± 4.48\% and 96.44 ± 0.96 \% respectively. 
As summarized in Table \ref{tab:similar_works_table} the deep learning model used in our study achieved the best results (97.8\% correlation for kinematics and 97.2 \% for grip force) compared to similar studies. Previous studies, only focused on wrist angle and grip forces neglecting the full range of human hand joint movements, however, our method successfully estimated 21 DoFs simultaneously. 
\iffalse
\begin{table*}[ht]
  \centering
  \caption{Average PCC scores for kinematics and force estimation across different methods }
  \begin{tabular}{cccccc}
    \toprule
    \multirow{2}{*}{Condition} & \multicolumn{2}{c}{Li et al. \cite{Li}} & \multicolumn{2}{c}{Mao et al. \cite{Mao}} & {Proposed method}\\
    \cmidrule(lr){2-3} \cmidrule(lr){4-5}
    & \multicolumn{1}{C{1.5cm}}{GCN+LSTM} & \multicolumn{1}{C{1.5cm}}{LSTM} & \multicolumn{1}{C{1.5cm}}{EMG} & \multicolumn{1}{C{1.5cm}}{EMG+ACC} \\
    \midrule
     Wrist angle/joint positions & 89.6& 71.5  & 86.41 &  96.44 & \textbf{97.8} \\
     Grip force & 91.9&  88.7  & 92.42 & 95.32 & \textbf{97.2}\\

    \bottomrule
  \end{tabular}
  
  \label{tab:similar_works_table}
\end{table*}

\fi
\begin{table*}[ht]
  \centering
  \caption{Average PCC scores for kinematics and force estimation across different methods}
  \begin{tabular}{cccccc}
    \toprule%\multirow{2}{*}{Condition} & 
    
    & \multicolumn{2}{c}{Li et al. \cite{Li}} & \multicolumn{2}{c}{Mao et al. \cite{Mao}} & {Proposed method}\\
    \cmidrule(lr){2-3} \cmidrule(lr){4-5} \cmidrule(lr){6-6}

    \multirow{1}{*}{method}& \multicolumn{1}{C{2cm}}{GCN+LSTM} & \multicolumn{1}{C{2cm}}{LSTM}& \multicolumn{2}{c}{SVR}   & {3DCNN + MLP}\\

    \cmidrule(lr){2-3} \cmidrule(lr){4-5} \cmidrule(lr){6-6}

    \multirow{1}{*}{input}& \multicolumn{2}{c}{EMG}  & \multicolumn{1}{C{1.5cm}}{EMG} & \multicolumn{1}{C{2cm}}{EMG + ACC} & {EMG}\\
    
    \cmidrule(lr){2-3} \cmidrule(lr){4-5} \cmidrule(lr){6-6}
    
    \multirow{1}{*}{channels no.}& \multicolumn{2}{c}{192 channels}  & \multicolumn{1}{C{1.5cm}}{7 channels} & \multicolumn{1}{C{2cm}}{14 channels} & {320 channels}\\
    \cmidrule(lr){2-3} \cmidrule(lr){4-5} \cmidrule(lr){6-6}
    
   \multirow{1}{*}{DoFs no.} & \multicolumn{2}{c}{2 DoFs } & \multicolumn{2}{c}{4 DoFs} & {21 DoFs }\\
   & \multicolumn{2}{c}{(1 Wrist angle + 1 Force)} & \multicolumn{2}{c}{(3 Wrist angles + 1 Force)} & {(20 hand joint positions + 1 Force)}\\
    
    \cmidrule(lr){2-3} \cmidrule(lr){4-5} \cmidrule(lr){6-6}
  \multirow{1}{*}{Force Level} & \multicolumn{2}{c}{30\% and 60\% MVC} & \multicolumn{2}{c}{under 30\% MVC} & {10\% and 30\% MVC }\\
    
    \midrule
     Wrist angle/joint positions & 89.6& 71.5  & 86.41 &  96.44 & \textbf{97.8} \\
     Grip force & 91.9&  88.7  & 92.42 & 95.32 & \textbf{97.2}\\

    \bottomrule
  \end{tabular}
  
  \label{tab:similar_works_table}
\end{table*}

\subsection{Force level effect}

\begin{figure}[htbp]
\vspace{-\baselineskip}
  \centering
  \includegraphics[width=0.48\textwidth]{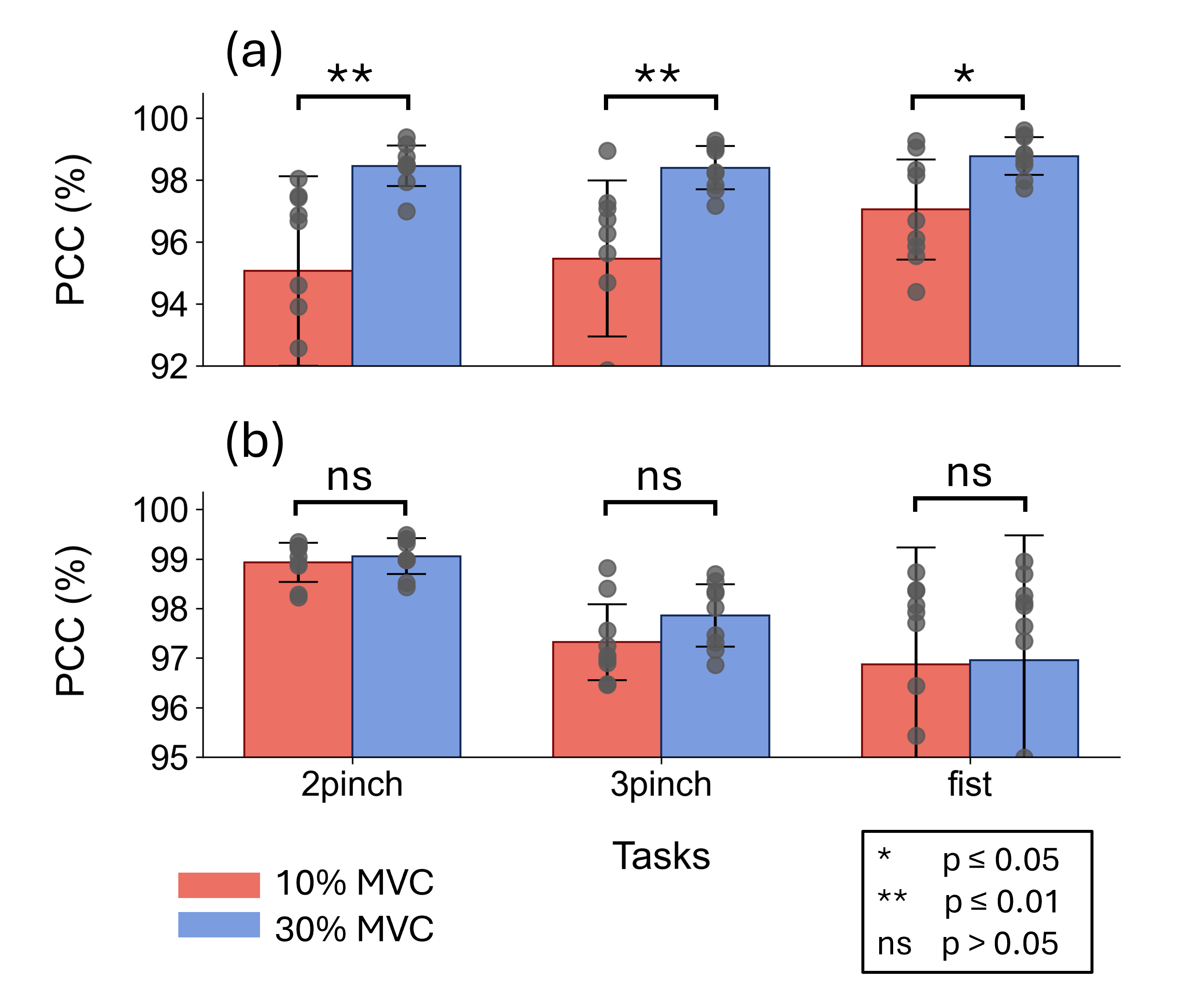}
  \caption{Model performance across different force levels.  \hyperref[fig:a]{(a)}   Force estimation performance for different tasks across different Force levels. Student's t-test statistical analysis is done to compare the groups. There is a significant difference between the predictions for 10\% and 30\% of MVC.  \hyperref[fig:b]{(b)}  Kinematics prediction results across different force levels. There is no significant difference between the model performance for kinematics across different force values.} 
  %\vspace{-\baselineskip}
  \label{fig:force_level_effect}
  \label{fig:a}
  \label{fig:b}
\end{figure}

The proposed model's performance is evaluated at different force levels. Figure \ref{fig:force_level_effect} shows the model performance for 10\% and 30\% MVC. The results are averaged over all subjects. These results indicate that the performance of the model for high and low forces is significantly different for all three movements (Fig \hyperref[fig:force_level_effect]{ \ref*{fig:force_level_effect}a}). 
The kinematics prediction accuracy is also evaluated across different force levels revealing no significant difference between the groups (Fig \hyperref[fig:force_level_effect]{ \ref*{fig:force_level_effect}b}). These findings suggest that higher force levels did not affect the model performance in kinematics estimation. However, they also imply that while the model can predict kinematics across varying force levels, it may have limitations in terms of accurately predicting higher force levels. This limitation could be due to the complex sEMG-Force nonlinear relation at higher force levels. As the force increases, the relationship becomes more complex due to the intricacies of motor unit recruitment and firing rates \cite{Didier}. This non-linearity might affect the model's ability to accurately predict force at higher levels.

\subsection{Limitations and future work}

This study has several limitations. The number of hand movements was restricted to three grasp types (2pinch, 3pinch, and fist), thereby limiting the generalization of the results to new and untested grip types and movements. Future experiments should include a broader range of hand movements. The lack of amputees is another limitation of this study which affects the study’s applicability to real clinical scenarios, as amputees are the main users of prosthetics. Future research should consider amputees to evaluate the model’s performance in practical scenarios. The real-time kinematics prediction was previously studied in \cite{Simpetru2} showing the model's ability in 20 joint position estimation, in this paper, the adapted model’s performance was evaluated in real-time force estimation, showing accurate online force predictions. However, the model currently has limitations in performing simultaneous kinetics and kinematics real-time predictions, which require further investigation.

\section{Conclusion}
In this paper, a method combining HD-EMG signals and deep learning techniques was presented to estimate 20 hand joint positions and grip forces simultaneously during three hand movements (2pinch, 3pinch, and fist closing and opening). The study involved 9 individuals performing three dynamic hand movements while applying forces at two different force levels (10\% and 30\% MVC). Given the limited research on simultaneous multi-DoF kinematics and kinetics estimation, our approach demonstrated superior performance in estimating 21 DoFs (20 hand joint positions and one force), facilitating comprehensive control of the human hand. Additionally, forces were examined with real-time resolution, aiming to enhance the model's effectiveness for real-time applications.

\section*{Acknowledgement}

The authors would like to acknowledge the scientific support and HPC resources provided by Friedrich-Alexander University. The hardware is funded by the German Research Foundation (DFG). Figure \hyperref[fig: overview]{ \ref*{fig: overview}a} was created with BioRender.com, which we gratefully acknowledge.

\section*{References}
\vspace*{-1.5em} 
%\bibliographystyle{ieeetr} 
%\bibliographystyle{unsrt} 
%\bibliography{references}

\end{document}